\documentstyle[12pt]{article}
     
\def\l{\lambda}
\def\D{{\cal D}}

\def\H{{\cal H}}
\def\E{{\rm E}\hskip-.55em{\rm I}}
\def\F{{\rm F}\hskip-.53em{\rm I}\,}
\def\ir{{\rm I}\hskip-.2em{\rm R}}
\def\half{\textstyle{\frac{1}{2}}}

\def\ra{\rightarrow}
\def\tint{{\textstyle{\int}}}

\def\b{\begin{eqnarray*}}     
\def\e{\end{eqnarray*}}       
\def\bn{\begin{eqnarray}}     
\def\en{\end{eqnarray}}       
\def\<{\langle}
\def\>{\rangle}

\def\{{\lbrace}
\def\}{\rbrace}
\title{New Measures for the \\Quantization of Systems with 
Constraints\footnote{Based on a presentation at the Inauguration
 Conference of 
the Asia Pacific Center for Theoretical Physics, Seoul, Korea, June 4-10, 
1996.\vskip.3cm \noindent BUTP-96/19 }}
\author{John R. Klauder\\
Departments of Physics and Mathematics\\
University of Florida\\
Gainesville, Fl  32611\\
and\\
Institut f\"ur Theoretische Physik\\
Universit\"at Bern\\
Bern, CH-3012}
\begin{document}
\maketitle
\begin{abstract}
Based on the results of a recent reexamination of the quantization of 
systems with first-class and second-class constraints from the point of view 
of coherent-state 
phase-space path integration, we give additional examples of the 
quantization procedure for reparameterization invariant Hamiltonians, for 
systems for which the original set of Lagrange  multipliers are elevated to 
dynamical variables, as well as extend the formalism to include cases of 
first-class constraints the operator form of which have a spectral gap about 
the value zero that characterizes the quantum constraint subspace. 
\end{abstract}
\section{Introduction}
This paper presents several examples of quantization for dynamical systems 
possessing closed first-class constraints following general procedures 
established in Ref.~\cite{kla}. For the reader's convenience we summarize 
several of the main ideas briefly; for a fuller discussion we suggest the 
reader consult \cite{kla}.

A classical action (summation implied)
  \bn I=\tint[p_j{\dot q}^j-H(p,q)-\l^a\phi_a(p,q)]\,dt\;,  \en
$1\leq j\leq J$, $1\leq a\leq A\leq 2J$, is said to describe a system with 
closed first-class constraints provided $\{p_j,q^j\}$ are dynamical 
variables, $\{\l^a\}$ are Lagrange multipliers, and
\bn &&\{\phi_a(p,q),\phi_b(p,q)\}=c_{ab}^{\;\;\;\;c}\,\phi_c(p,q)\;,\nonumber\\
  &&\{\phi_a(p,q),H(p,q)\}=h_a^{\;\;b}\,\phi_b(p,q)\;,  \en
where $\{\cdot,\cdot\}$ denotes the Poisson bracket.

In the quantum theory we assume that ($\hbar=1$)
\bn &&[\Phi_a(P,Q),\Phi_b(P,Q)]=ic_{ab}^{\;\;\;\;c}\,\Phi_c(P,Q)\;,\nonumber\\
  &&[\Phi_a(P,Q),{\cal H}(P,Q)]=ih_a^{\;\;b}\,\Phi_b(P,Q)\;,  \en
where $\Phi_a$ and $\H$ denote self-adjoint constraint and Hamiltonian 
operators, respectively. In \cite{kla} a coherent-state path integral was 
defined (by a suitable lattice limit) in such a way that
\bn &&\hskip-1cm{\cal M}\int\exp\{i\tint_0^T[p_j
{\dot q}^j-H(p,q)-\l^a\phi_a(p,q)]\,dt\}\,\D p\,\D q\,\D C(\l)\nonumber\\
&&\hskip.3cm=\<p'',q''|e^{-i\H T}\E\,|p',q'\>[1+O(\delta)]\nonumber\\
&&\hskip.3cm=\<p'',q''|\E\,e^{-i\H T}\E\,|p',q'\>[1+O(\delta)]\;. \en
In this expression $\cal M$ is a formal normalization and $C(\l)$, $\tint\D 
C(\l)=1$, is a measure on the Lagrange multipliers designed to introduce (at 
least) one factor of a suitable {\it projection operator} $\E\,$. Because 
the set of self-adjoint operators $\{\Phi_a\}$ forms a closed Lie algebra, 
we may define $\exp(-i\xi^a\Phi_a)$, for appropriate sets $\{\xi^a\}$ of 
real parameters, as unitary group operators. For a {\it compact group} with 
a normalized Haar measure $\delta\xi$, $\tint\delta\xi=1$, then
  \bn \E\,=\tint e^{-i\xi^a\Phi_a}\,\delta\xi= \E\,(\Phi_a=0)\;.  \en
For a {\it noncompact group}, some generators of which have continuous 
spectra, then we choose a weight function $f(\xi)$ such that
  \bn  \E\,=\tint e^{-i\xi^a\Phi_a}\,f(\xi)\,\delta\xi=\E\,(\Sigma 
\Phi_a^2\leq\delta^2)\;,  \en
for $0<\delta\ll 1$. [Effectively, for a compact group, we may choose 
$\delta=0$, and in that case the term $O(\delta)\equiv0$ in both lines of 
(4).] As discussed in \cite{kla} the properly defined path integral with a 
suitable integration measure for the Lagrange multipliers---essentially 
different than in the Faddeev treatment \cite{fad}---leads, automatically, 
to gauge invariant results for compact groups, and after a suitable 
$\delta$-limiting process $\delta\ra0$ (see below) to gauge invariant 
results for noncompact groups. Moreover, no $\delta$-functionals of the 
classical constraints nor $\delta$-functionals of any subsidary conditions 
are introduced, and as a consequence no Faddeev-Popov determinant appears. 
Ambiguities that often may arise with such determinants \cite{gri} are 
thereby avoided at the outset.

In this paper we illustrate such quantization procedures for the example of 
a reparameterization invariant Hamiltonian which was not considered in 
\cite{kla} (see also \cite{gov}). We also consider the situation in which an 
original set of Lagrange multipliers are themselves elevated to the status 
of dynamical variables and used to define an extended dynamical system which 
is completed with the addition of suitable conjugates and new sets of 
constraints and their associated Lagrange multipliers. Finally, we extend 
the formalism to include rather general phase-space constraints the operator 
form of which has a spectral gap about zero. We generally follow the 
notation of \cite{kla}.
\section{Reparameterization Invariant Dynamics}
Let us start with a single degree of freedom $(J=1)$ and the action
 \bn \tint[p{\dot q}-H(p,q)]\,dt\;.  \en
We next promote the independent variable $t$ to a dynamical variable, 
introduce $s$ as its conjugate momentum (often called $p_t$), enforce the 
constraint $s+H(p,q)=0$, and lastly introduce $\tau$ as a new independent 
variable. This modification is realized by means of the classical action
\bn \tint\{pq'+st'-\l[s+H(p,q)]\}\,d\tau\;,  \en
where $q'=dq/d\tau$, $t'=dt/d\tau$, and $\l=\l(\tau)$ is a Lagrange 
multiplier. The coherent-state path integral is constructed so that  
\bn &&\hskip-.5cm
{\cal M}\int\exp(\!\!(i\tint\{pq'+st'-\l[s+H(p,q)]\}\,dt)\!\!)\,\D p\,
\D q\,\D s\,\D t\,\D C(\l)\nonumber\\
&&\hskip1.5cm=\<p'',q'',s'',t''|\E\,|p',q',s',t'\>\;,  \en
where
  \bn  &&\E\,\equiv \int_{-\infty}^\infty 
e^{-i\xi[S+\H(P,Q)]}\;\frac{\sin(\delta\xi)}{\pi\xi}\,d\xi\nonumber\\
&&\hskip.4cm=\E\,(-\delta<S+\H(P,Q)<\delta)\;.  \en
The result in (9) and (10) represents as far as we can go without choosing 
${\cal H}(P,Q)$.

To gain further insight into such expressions we specialize to the case of 
the free particle, $\H=P^2/2$. Then it follows \cite{kla} that
\bn  &&\<p'',q'',s'',t''|\E\,|p',q',s',t'\>\nonumber\\    &&\hskip.1cm=\pi^{-1}
\int_{-\infty}^\infty\exp[-\half(k-p'')^2-\half(\half 
k^2+s'')^2\nonumber\\
  &&\hskip2.85cm+ik(q''-q')-i\half k^2(t''-t')\nonumber\\
  &&\hskip2.85cm-\half(k-p')^2-\half(\half k^2+s')^2]\,dk\nonumber\\  &&\hskip2.3cm\times\frac{2\sin[\delta(t''-t')]}{(t''-t')}+O(\delta^2)\;.\en
For any $\delta$ such that $0<\delta\ll1$, we observe that this expression represents 
a {\it reproducing kernel} which in turn defines an associated {\it 
reproducing kernel Hilbert space} composed of bounded, continuous functions 
given, for arbitrary complex numbers $\{\alpha_k\}$, phase-space points 
$\{p_k,q_k,s_k,t_k\}$, and $K<\infty$, by
\bn  \psi(p,q,s,t)=\sum_{k=0}^K\alpha_k\<p,q,s,t|\E\,|p_k,q_k,s_k,t_k\>\;,  
\en
or as the limit of Cauchy sequences of such functions in the norm defined by 
means of the inner product given by
\bn (\psi,\psi)=\tint|\psi(p,q,s,t)|^2\,dp\,dq\,ds\,dt/(2\pi)^2  \en
integrated over $\ir^4$. 

Let us next consider the {\it reduction} \cite{kla} of the reproducing 
kernel given by 
\bn &&\hskip-1cm\<p'',q'',t''|p',q',t'\>\nonumber\\  
&&\equiv\lim_{\delta\ra0}\frac{1}{4\sqrt{\pi}\delta}\int\<p'',q'',s'',t''|
\E\,|p',q',s',t'\>\,ds''\,ds'\nonumber\\
&&=\pi^{-1/2}\int\exp[-\half(k-p'')^2-\half(k-p')^2\nonumber \\
  &&\hskip2.68cm+ik(q''-q')-i\half k^2(t''-t')]\,dk\;,  \en
which in turn generates a {\it new} reproducing kernel in the indicated 
variables. For the resultant kernel it is straightforward to demonstrate, 
for any $t$, that 
 \bn \int\<p'',q'',t''|p,q,t\>\<p,q,t|p',q',t'\>\,dp\,dq/(2\pi) 
=\<p'',q'',t''|p',q',t'\> \;. \en
This relation implies that the span of the vectors 
$\{|p,q\>\equiv|p,q,0\>\}$ is identical with the span of the vectors 
$\{|p,q,t\>\}$, meaning further
that the states $\{|p,q,t\>\}$ form a set of {\it extended coherent states}, 
which are extended with respect to $t$ in the sense of Ref.~\cite{kwh}. 
Observe how the time variable has become distinguished by the criterion 
(15). Consequently, we may properly interpret
 \bn 
\<p'',q'',t''|p',q',t'\>\equiv\<p'',q''|e^{-i(P^2/2)(t''-t')}|p',q'\>\;, \en
namely as the conventional, single degree of freedom, coherent state matrix 
element of the evolution operator appropriate to the free particle. 

To further demonstrate this interpretation as the dynamics of the free 
particle, we may pass to sharp $q$ matrix elements with the observation that
\bn &&\hskip-1.5cm\<q''|e^{-i(P^2/2)(t''-t')}|q'\>\nonumber\\
&&\equiv\frac{\pi^{1/2}}{(2\pi)^2}\int\<p'',q''|e^{-i(P^2/2)(t''-t')}|p',q'
\>\,dp''\,dp'\nonumber\\
&&=\frac{1}{2\pi}\int\exp[\,ik(q''-q')-i\half k^2(t''-t')]\,dk\nonumber\\
&&=\frac{\;\;e^{i(q''-q')^2/2(t''-t')}}{\sqrt{2\pi i(t''-t')}}\;, \en
which is evidently the usual result.
\section{Elevating the Lagrange Multiplier to an Additional Dynamical 
Variable}
Sometimes it is useful to consider an alternative formulation of a system 
with constraints in which the initial Lagrange multipliers are regarded as 
dynamical variables, complete with conjugate variables, and to introduce new 
constraints as needed. For example, let us start with a single degree of 
freedom system with a single first-class constraint specified by the action 
functional
  \bn \tint[p{\dot q}-H(p,q)-\l\phi(p,q)]\,dt\;,  \en
where $\phi(p,q)$ represents the constraint and $\l$ the Lagrange 
multiplier. Instead, let us replace this action functional by
  \bn \tint[p{\dot q}+\pi{\dot\l}-H(p,q)-\xi\pi-\theta\phi(p,q)]\,dt\;. \en
In this expression we have introduced $\pi$ as the canonical conjugate to 
$\l$, the Lagrange multiplier $\xi$ to enforce the constraint $\pi=0$, and 
the Lagrange multiplier $\theta$ to enforce the original constraint 
$\phi=0$. Observe that $\{\pi,\phi(p,q)\}=0$, and therefore the constraints 
remain first class in the new form. The path integral expression for the 
extended form reads
  \bn &&\hskip-.7cm{\cal M}\int\exp\{i\tint[p
{\dot q}+\pi{\dot\l}-H(p,q)-\xi\pi-\theta\phi(p,q)]\,dt\}\,\D p\,
\D q\,\D\pi\,\D\l\,\D C(\xi,\theta)\nonumber\\
&&\hskip1.3cm=\<p'',q'',\pi'',\l''|e^{-i{\cal H}T}\E\,|p',q',\pi',\l'\>\;. 
\en
In this expression
  \bn  \E\,=\E\,(-\delta<\Phi(P,Q)<\delta)\,\E\,(-\delta<\Pi<\delta)\;. \en
Consequently, the complete propagator factors into two terms,
 \bn &&\<p'',q'',\pi'',\l''|e^{-i{\cal H}T}\E\,|p',q',\pi',\l'\>\nonumber\\
&&\hskip2cm=\<p'',q''|e^{-i{\cal H}T}\E\,(-\delta<\Phi(P,Q)<\delta)|p',q'\>
\nonumber\\
&&\hskip2.15cm\times\<\pi'',\l''|\E\,(-\delta<\Pi<\delta)|\pi',\l'\>\;. \en
The first factor is exactly what would be found by the appropriate path 
integral of the original classical system with only the single constraint 
$\phi(p,q)=0$ and the single Lagrange multiplier $\l$. The second factor 
represents the modification introduced by considering the extended system. 
Note however that with a suitable $\delta$-limit \cite{kla} the second 
factor reduces to a product of terms, one depending on the ``$\;''\;$'' 
arguments, the other depending on the ``$\;'\;$'' arguments. This result for 
the second factor implies that it has become the reproducing kernel for a 
{\it one-dimensional} Hilbert space, and when multiplied by the first factor 
it may be ignored entirely. In this way it is found that the quantization of 
the original and extended systems leads to identical results.
\section{Constraints with a Spectral Gap about Zero}
Let us initially consider the example of a classical constraint of the form
  \bn  \phi(p,q)=p^2+q^2+q^4-c=0  \en
for a suitable value of $c$.
Here the natural operator form of the constraint has a discrete spectrum, 
but a spectrum without any particular regularity (in contrast with the case 
where the quartic term is absent). The integral in (6) does not define the 
desired projection operator for any locally integrable $f$. Thus we must 
entertain the idea that the integral designed to yield a projection operator 
may in fact not always do so, and consequently, we need to develop a 
formulation in which the relevant integral yields the desired projection 
operator {\it plus an arbitrarily small correction term} that must be sent 
to zero in some subsequent limiting operation.  
 In particular, motivated by the case at hand, let us consider the 
expression
\bn  \F\equiv\tint e^{-i\xi\Phi}\,f(\xi)\,d\xi\;,  \en
where the spectrum of $\Phi$ is purely discrete, includes zero, and has a 
gap between the eigenvalue zero and the next closest eigenvalue. In 
particular, if $\Phi|m\>=\phi_{(m)}|m\>$ characterizes a complete set of 
eigenvectors and eigenvalues, then, excluding the particular value 
$\phi_{(m)}=0$, we set $\Delta\equiv\inf(|\phi_{(m)}|)$  
and assume that $\Delta>0$. It is immaterial whether or not there are 
degeneracies.  As noted, one of the eigenvalues $\phi_{(m)}$ is zero, and it 
is onto the subspace of that eigenvalue that we want to project. We let 
$\E\,$ denote the projection operator onto the desired subspace.
Clearly, if we choose the integration in (24) as an average over the 
interval from $[-L,L]$, $L>0$, then
\bn  \F=\E\,+W   \en
where $W$ is a bounded operator with a bound given by 
\bn  \|W\|\leq\frac{1}{L\Delta}\;.  \en
The operator $W$ is not a projection operator, but its operator norm can be 
made arbitrarily small by choosing $L$ large enough. In other words, 
$\F=\E\,$ apart from an operator of arbitrarily small norm. 

For the case of closed first-class constraints, only a single projection 
operator is needed to effect the projection of the system onto the physical 
Hilbert space. In this case we may easily generalize the discussion of 
\cite{kla} to arrive at the conclusion that there exists a measure $L(\xi)$, 
$\tint dL(\xi)=1$,  on the Lagrange multipliers at the initial time such 
that, heuristically,
\bn  &&\hskip-.5cm{\cal M}\int\exp\{i\tint[p_j
{\dot q}^j-H(p,q)-\l^a\phi_a(p,q)]\,dt-i\xi^a\phi_a(p',q')\}\,\D p\,\D q\,
d L(\xi)\nonumber\\
 &&\hskip1cm=\<p'',q''|e^{-i\H T}\E\,|p',q'\>[1+O(L^{-1})]\;.  \en
Here the measure $L(\xi)$---and thereby the projection operator $\E\,$---are 
adequately defined by the requirement that
\bn \F\equiv\tint e^{-i\xi^a\Phi_a}\,d L(\xi)\equiv \E\,+W\;,  \en
where $\E\,$ represents the projection operator onto the appropriate quantum 
constraint subspace and $W$ is an operator such that $\|W\|\leq 
(L\Delta)^{-1}$, for some $\Delta>0$. Just as was the case for the 
$\delta$-limiting procedure, a subsequent limit as $L\ra\infty$ can be 
introduced to the reproducing kernel, leading, in general, to a reduction of 
the appropriate reproducing kernel in the sense of \cite{kla}. On the other 
hand, observe that the 
original path integral (27) (with $L<\infty$) defines a reproducing kernel 
with an inner product having a local integral representation exactly as for 
the unconstrained system, and furthermore, if $L$ is huge, e.g., 
$L\Delta=10^{50}$, then the errors represented by $W$ are negligible.
\subsubsection*{Constraints with mixed spectra}
Although we have concentrated on constraint operators with purely discrete 
spectra, all that was really essential in showing that $\F-\E\,$ was an 
operator with a small norm was the existence of a gap in the spectrum at 
zero which we have termed $\Delta$ with $\Delta>0$. So long as this gap 
remains it is possible to argue that $\F-\E\,$ remains small even though the 
constraint operators may have continuous spectra away from zero. Thus 
without further comment we may extend the results of this section to include 
such cases. For example, in this extension one may consider the classical 
constraint given by $p^2-1/(1+q^2)^{3/2}+q^2/(1+q^2)=0$ which in its quantum 
form has a normalizable eigenstate that satisfies the quantum constraint 
condition, but also has a continuous spectrum separated by an appropriate 
gap.
\section*{Acknowledgements}
Thanks are expressed to J. Govaerts and S. Shabanov for their interest in 
the present formulation of the quantization of constrained systems.

\end{document}